\documentclass{article}
\usepackage{jim,amsmath}
\usepackage[utf8]{inputenc}
\usepackage[french]{babel}
\usepackage[T1]{fontenc}

\usepackage[table]{xcolor}

\usepackage{tikz}
\usetikzlibrary{arrows,decorations.markings}
\usetikzlibrary{cd}
\usetikzlibrary{shapes.geometric,fit}
\usetikzlibrary{positioning}
\tikzcdset{every label/.append style = {font = \tiny}}

\usepackage{leadsheets}

\usepackage{mathtools}
\usepackage{amsmath}
\usepackage{amsfonts}
\usepackage{amsthm}
\usepackage{amssymb}

\usepackage{graphicx}
\usepackage{wrapfig}
\usepackage{float}

\usepackage{hyperref}
\usepackage{subcaption}

\usepackage{csquotes}
\usepackage{comment}

\usepackage{tabularx}
\newcolumntype{Y}{>{\centering\arraybackslash}X}
\newcolumntype{Z}{>{\raggedright\arraybackslash}X}
\definecolor{tableShade}{gray}{0.9}

\usepackage[shortlabels]{enumitem}

\usepackage{relsize}

\usepackage{hyperref}

\usepackage{float}

\newcommand{\LSMod}{\texttt{Mod} }
\newcommand{\LSpp}{\texttt{++} }
\newcommand{\LSmm}{\texttt{-\hspace{0.5pt}-} }
\newcommand{\LSI}{\writechord{I}}
\newcommand{\LSII}{\writechord{II}}
\newcommand{\LSIII}{\writechord{III}}
\newcommand{\LSIV}{\writechord{IV}}
\newcommand{\LSV}{\writechord{V}}
\newcommand{\LSVI}{\writechord{VI}}
\newcommand{\LSVII}{\writechord{VII}}
\newcommand{\LSi}{\writechord{i}}
\newcommand{\LSii}{\writechord{ii}}
\newcommand{\LSiii}{\writechord{iii}}
\newcommand{\LSiv}{\writechord{iv}}
\newcommand{\LSv}{\writechord{v}}
\newcommand{\LSvi}{\writechord{vi}}
\newcommand{\LSvii}{\writechord{vii}}
\newcommand{\LStwo}{\texttt{2}}
\newcommand{\LSthree}{\texttt{3}}
\newcommand{\LSfour}{\texttt{4}}
\newcommand{\LSMm}{\texttt{M}$\leftrightarrow$\texttt{m} }

\DeclareRobustCommand{\svdots}{
  \vcenter{%
    \offinterlineskip
    \hbox{.}
    \vskip0.25\normalbaselineskip
    \hbox{.}
    \vskip0.25\normalbaselineskip
    \hbox{.}%
  }%
}

\title{LiveScaler : Contrôler en live l'harmonie d'un morceau de musique électronique}
\oneauthor
 {Alice Rixte} {Université de Bordeaux \\ alice.rixte@u-bordeaux.fr }

\begin{document}

\maketitle

\sloppy

\begin{abstract}
Dans le contexte de l'Electronic Dance Music (EDM), de nombreux artistes utilisent des techniques de \emph{DJing} pour performer leurs propres productions sur scène, se privant ainsi de l'accès à la structure interne de leurs morceaux, et en particulier de l'équivalent de leur partition : les fichiers MIDI joués par des instruments virtuels. De plus, si l'artiste remixe ou interprète sa propre production, le nombre de pistes pouvant être contrôlées simultanément est limité sans un outil adapté.

Cet article présente le logiciel LiveScaler, qui permet de contrôler en live l'harmonie et la hauteur des notes de tous les instruments virtuels d'un morceau de musique électronique. Un ensemble restreint de transformations de l'espace des hauteurs de notes, les \emph{transformations affines}, est introduit. Ces transformations sont appliquées à tous les flux MIDI d'un morceau composé préalablement. Une implémentation utilisant Max MSP en combinaison avec Ableton Live (Max for Live) est proposée. Une attention particulière est portée aux questions de contrôle, de \emph{mapping} et de mise en pratique dans le cadre de l'EDM.
\end{abstract}

\section{Introduction}
Aux origines de la \emph{club culture}, le rôle des DJs était surtout de jouer des enregistrements musicaux pour un public, en général dansant. Avec l’évolution rapide de la technologie et en particulier l’omniprésence de l’audionumérique à partir des années 2000, de nombreux DJs sont devenus à la fois compositeurs, producteurs et performeurs de leurs propres morceaux. C'est notamment le cas des artistes évoluant dans le milieu de l'Electronic Dance Music (EDM)\footnote{\emph{Electronic Dance Music} (EDM) est un terme parapluie regroupant de nombreux genres de musique électronique tels que la house, la techno, la trance, la drum n bass, le dubstep, etc. }.

Ces artistes composent le plus souvent leur musique à l’aide d’une station audionumérique (Digital Audio Workstation ou DAW) qui leur permet de combiner sampleurs, synthétiseurs et enregistrements pour créer un morceau de musique complet. Recréer en live un tel morceau, composé souvent de plusieurs dizaines de pistes, reste difficile et conditionné par les contrôles proposés par les DAW. Pour leur performance live, la plupart des artistes vont ainsi choisir entre interpréter certaines pistes spécifiques à l'aide de synthétiseurs ou sampleurs, remixer en live un morceau préparé au préalable ou utiliser des techniques de DJing leur permettant de mixer entre eux des rendus audios en leur appliquant divers effets \cite{ferreira2008sound},\cite{magana2018performance}.  Dans ces contextes,  modifier le rythme ou l'harmonie d’un morceau joué en live est difficilement réalisable dans ces contextes. En effet, cela requerrait un accès à la structure interne du morceau, ce que le DJing ne permet pas, ou bien de contrôler toutes les pistes simultanément, ce qui est précisément ce que nous proposons de faire ici.

Cet article présente LiveScaler, qui per\-met de modifier en live la structure harmonique d'un morceau de musique électronique\footnote{Une vidéo de démonstration est disponible à l'adresse suivante : \href{ https://youtu.be/Cn0HBgWS5Pw}{youtu.be/Cn0HBgWS5Pw}}. En partant d'un morceau composé au préalable, LiveScaler permet d'appliquer des transformations MIDI à l'ensemble des instruments virtuels. LiveScaler conserve l'ensemble des caractéristiques du morceau initial mais agit sur la hauteur des notes, permettant de changer en live l'harmonie du morceau ou de générer de nouvelles mélodies. En particulier, l'accent est mis sur la possibilité d'utiliser LiveScaler dans un DAW, ici Ableton Live\footnote{Une implémentation de LiveScaler avec Max For Live est disponible en libre accès sur Github : \href{https://github.com/autonym8/LiveScaler}{github.com/autonym8/LiveScaler}}, tout en minimisant les contraintes et les connaissances techniques nécessaires pour son utilisation. Dans Ableton Live, LiveScaler peut s'utiliser aussi bien dans le mode \emph{session} que dans le mode \emph{arrangement}, et s'ajoute à toutes les pistes MIDI que l'on souhaite être impactées par les transformations.

Cet article est organisé de la manière suivante : nous commencerons par définir les transformations affines, un ensemble restreint de transformations de l'espace des hauteurs de notes. Ensuite, nous présenterons Live\-Scaler, qui implémente l'application de ces transformations en live à un nombre arbitraire d'instruments virtuels. Enfin, nous décrirons la manière dont nous avons utilisé LiveScaler dans le cadre d'une performance live d'EDM . Nous terminerons par une comparaison de LiveScaler avec les travaux et outils existants.
\section{Transformations de l'espace des hauteurs}
Dans cette section nous présentons les transformations de l'espace des hauteurs que nous utilisons dans LiveScaler. L'idée principale est d'associer à chaque note une nouvelle note, qui sera jouée en lieu et place de la note initiale par l'ensemble des instruments du morceau de musique joué. LiveScaler permet d'appliquer deux types de transformations. D'une part les \emph{transformations périodiques sur un intervalle}, que nous présenterons après avoir défini l'espace des hauteurs de note sur lequel agissent nos transformations. D'autre part les \emph{transformations affines}, dont nous décrirons les propriétés musicales. Enfin, nous terminerons en restreignant l'écart de hauteur entre la note initial et la note transformée afin de résoudre de potentiels problèmes de tessiture.

\subsection{Espace des hauteurs}
Dans cet article, nous nous plaçons dans l'espace des hauteurs linéaire. Celui-ci est obtenu à partir de la demi-droite réelle positive des fréquences, en fixant une origine arbitraire de fréquence $\alpha$ (par exemple $440$ Hz). Ici, nous appellerons cette fréquence l'\emph{ancre}.  À partir d'une fréquence $f>0$, on détermine une hauteur $p$ par

$$p = \log_2(f/\alpha)$$

On obtient alors la droite réelle : $p\in \mathbb{R}$. En discrétisant cette droite, on peut obtenir une note $n$ à partir de la hauteur $p$ par
$$n = \lfloor \beta p \rfloor$$
\noindent avec $\beta\in \mathbb{N}^*$ le nombre de divisions de l'octave. On obtient ainsi le tempérament à division multiple $\beta$-TET \footnote{TET signifie Tone Equal Temperament en anglais}. En particulier, lorsque $\beta = 12$ et $\alpha = 8.1758$ Hz \footnote{ $8.1758$ à $0$ est la fréquence (ici arrondie) correpondant à la note \writechord{C}$_{-1}$ et la note MIDI $0$.}, on retrouve le codage des hauteurs MIDI.

Ainsi, pour définir un tempérament multiple, nous avons besoins de deux paramètres : l'\emph{ancre} $\alpha$  et le nombre de divisions de l'octave $\beta$, que nous appelons \emph{base}. On définit alors \begin{align*}
  T\langle \alpha, \beta \rangle &: \mathbb{R}_+^* \rightarrow \mathbb{Z}\\
  &:f \mapsto \lfloor \beta \log_2(f/\alpha)\rfloor
\end{align*}

Dans cet article, nous nous intéresserons à l'\emph{espace des hauteurs linéaires} discrétisé $\mathbb{Z} = T\langle\alpha, \beta \rangle (\mathbb{R^*_+})$. Par abus de langage, on appellera ici la fonction $T\langle \alpha, \beta \rangle$ le \emph{tempérament égal} d'ancre $\alpha$ et de base $\beta$. En pratique, $\alpha$ correspondra toujours à une note MIDI, et on se permettra d'écrire $\alpha =$ \writechord{C}$_5$ pour indiquer que la fréquence correspondant à la note \writechord{C}$_5$ est associée à $0$ dans $\mathbb{Z}$.

Nous définissons alors une \emph{transformation de gamme} comme une fonction qui à toute hauteur de note associe une nouvelle hauteur de note a priori quelconque, autrement dit une fonction de $\mathbb{Z}$ dans $\mathbb{Z}$. Cette transformation est \emph{a priori} indépendante de la manière dont $\mathbb{Z}$ est relié à l'espace des fréquences.

\subsection{Transformations périodiques sur un intervalle}
\begin{figure}[htbp]
  \centering
  \begin{tikzpicture}[baseline= (a).base]

    \node[scale=1] (a) at (0,0){
      \begin{tikzcd}[column sep=0mm, minimum width = 0mm, minimum height=7mm, row sep=-0.05cm]
        \svdots   & \svdots & \hspace{20mm} & \svdots & \svdots \\
        \writechord{B}_{5}  & 11 & & 11 & \writechord{B}_{5}  \\
        \writechord{A#}_{5} & 10 & & 10 & \writechord{A#}_{5} \\
        \writechord{A}_{5}  & 9  & & 9 & \writechord{A}_{5}   \\
        \writechord{G#}_{5} & 8  & & 8 & \writechord{G#}_{5}  \\
        \writechord{G}_{5}  & 7  & & 7  & \writechord{G}_{5} \\
        \writechord{F#}_{5} & 6  & & 6  & \writechord{F#}_{5} \\
        \writechord{F}_{5}  & 5  & & 5  & \writechord{F}_{5}  \\
        \writechord{E}_{5}  & 4  & & 4  & \writechord{E}_{5}  \\
        \writechord{D#}_{5} & 3  & & 3  & \writechord{D#}_{5} \\
        \writechord{D}_{5}  & 2  & & 2  & \writechord{D}_{5}  \\
        \writechord{C#}_{5} & 1  & & 1  & \writechord{C#}_{5} \\
        \writechord{C}_{5}  & 0  & & 0  & \writechord{C}_{5}  \\
        \svdots         & \svdots & & \svdots &         \svdots
        \arrow[from=2-2, to=2-4]
        \arrow[from=3-2, to=2-4]
        \arrow[from=4-2, to=4-4]
        \arrow[from=5-2, to=6-4]
        \arrow[from=6-2, to=6-4]
        \arrow[from=7-2, to=8-4]
        \arrow[from=8-2, to=8-4]
        \arrow[from=9-2, to=9-4]
        \arrow[from=10-2, to=9-4]
        \arrow[from=11-2, to=11-4]
        \arrow[from=12-2, to=13-4]
        \arrow[from=13-2, to=13-4]
      \end{tikzcd}
    };
  \end{tikzpicture}
  \caption{La quantisation vers la gamme majeure est périodique sur l'octave : le même motif est répété sur chaque octave.}
  \label{fig:quantmaj}
\end{figure}
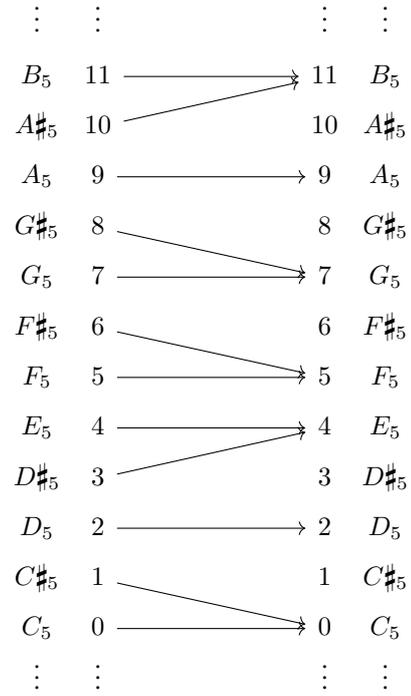

Nous définissons ici un ensemble de transformations des hauteurs facilement implémentable : les transformations périodiques sur un intervalle positif $n\in \mathbb{N}$. Ces transformations s'obtiennent en donnant l'image de toutes les notes se situant dans l'intervalle $i$ , typiquement une octave ($i = 12$), puis en répétant ce motif verticalement en additionnant (ou soustrayant) l'intervalle considéré. Plus précisément, pour une transformation $X : \mathbb{Z} \rightarrow \mathbb{Z}$, $X$ est périodique sur l'intervalle$i$ lorsque la fonction $Y : n \mapsto X(n) - n$ est périodique de période $i$. Autrement dit, $$Y(iq + n) = Y(n)$$
\noindent avec  $0 \leq n < i$ et tout $q\in \mathbb{Z}$.

On peut par exemple définir une transformation qui quantiser les $12$ demi-tons chromatique vers une gamme de notre choix, par exemple la gamme majeure dans la Figure \ref{fig:quantmaj}.

\subsection{Transformations affines}

Nous présentons ici les \emph{transformations affines}, c'est-à-dire les fonctions de la forme $A\langle\mu,\tau\rangle : n \mapsto \mu n + \tau$ avec $\mu$ le \emph{mode} de la transformation affine et $\tau$ sa \emph{transposition}. Nous allons classifier ces transformations en fonction de $mu$, qui détermine, en fonction du mode départ, le mode dans lequel on se trouvera après avoir appliqué la transformation.

Les transformations affines ont la propriété importante de préserver les classes de hauteur \footnote{En effet, pour toute base $\beta\in \mathbb{N}^*$, $\forall n_1,n_2 \in \mathbb{Z}, n_1 \equiv n_2 \mod \beta \implies \mu n_1 + \tau \equiv \mu n_2 + \tau \mod \beta$. Avec $\beta=12$, on obtient le résultat pour les classes de hauteurs dodécaphoniques. }  (au sens de Forte \cite{forte1973structure}) c'est-à-dire que si deux notes sont identiques à l'octave près, alors elles le seront toujours une fois la transformation affine appliquée.

La suite de cette section étudie plusieurs exemples afin de donner au lectorat un aperçu de leur expressivité. Dans l'ensemble de ces exemples, on se placera dans le tempérament égal $T\langle $\writechord{C}$_5,12\rangle$ : la note \writechord{C}$_5$ correspondra ainsi à l'entier $0$, \writechord{C#}$_5$ à $1$, \writechord{C}$_6$ à $12$, \writechord{B}$_4$ à $-1$, \writechord{C}$_4$ à $-12$, etc. Nous nous concentrerons ici sur les cas $\mu = 1$, $\mu = -1$ et $\mu = 2$.

\subsubsection{Transpositions}

\begin{figure}[htbp]
  \centering
  \begin{tikzpicture}[baseline= (a).base]

    \node[scale=1] (a) at (0,0){
      \begin{tikzcd}[column sep=0mm, minimum width = 7mm, minimum height=7mm, row sep=0cm]
        \svdots   & \svdots & \hspace{20mm} & \svdots & \svdots \\
        \writechord{E}_{5}  & 4  & & 4  & \writechord{E}_{5}  \\
        \writechord{D#}_{5} & 3  & & 3  & \writechord{D#}_{5} \\
        \writechord{D}_{5}  & 2  & & 2  & \writechord{D}_{5}  \\
        \writechord{C#}_{5} & 1  & & 1  & \writechord{C#}_{5} \\
        \writechord{C}_{5}  & 0  & & 0  & \writechord{C}_{5}  \\
        \writechord{B}_{4}  & -1 & & -1 & \writechord{B}_{4}  \\
        \writechord{A#}_{4} & -2 & & -2 & \writechord{A#}_{4} \\
        \writechord{A}_{4}  & -3 & & -3 & \writechord{A}_{4}  \\
        \svdots         & \svdots & & \svdots &         \svdots
        \arrow[from=3-2, to=1-4, dotted]
        \arrow[from=4-2, to=2-4]
        \arrow[from=5-2, to=3-4]
        \arrow[from=6-2, to=4-4]
        \arrow[from=7-2, to=5-4]
        \arrow[from=8-2, to=6-4]
        \arrow[from=9-2, to=7-4]
        \arrow[from=10-2, to=8-4, dotted]
      \end{tikzcd}
    };
  \end{tikzpicture}
  \caption{La transformation $A \langle 1,2 \rangle : n \mapsto n + 2$ correspond à la  transposition d'un ton vers l'aigu}
  \label{fig:transp}
\end{figure}
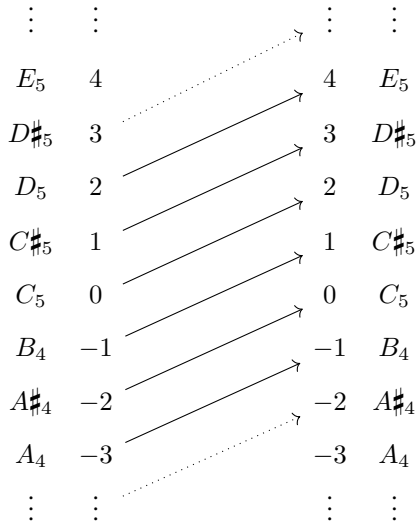
Lorsque $\mu = 1$, les transformations affines $A\langle 1,\tau \rangle : n \mapsto n + \tau$ permettent de représenter toutes les transpositions possibles (voir Figure \ref{fig:transp}).

\subsubsection{Inversions}

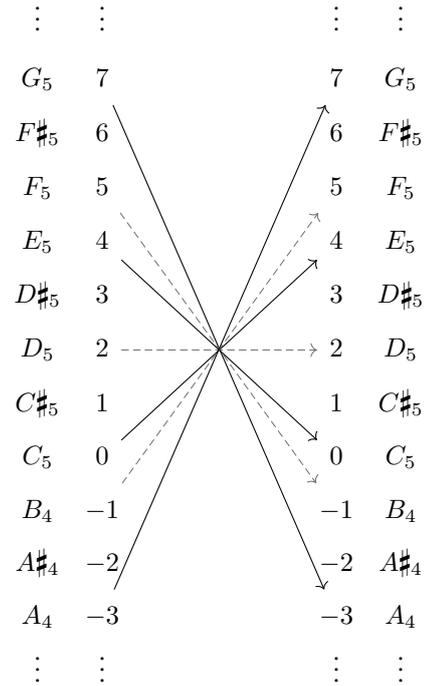
\begin{figure}[htbp]
  \centering
  \begin{tikzpicture}[baseline= (a).base]

    \node[scale=1] (a) at (0,0){
      \begin{tikzcd}[column sep=0mm, minimum width = 0mm, minimum height=7mm, row sep=0cm]
        \svdots   & \svdots & \hspace{20mm} & \svdots & \svdots \\
        \writechord{G}_{5}  & 7  & & 7  & \writechord{G}_{5}  \\
        \writechord{F#}_{5} & 6  & & 6  & \writechord{F#}_{5} \\
        \writechord{F}_{5}  & 5  & & 5  & \writechord{F}_{5}  \\
        \writechord{E}_{5}  & 4  & & 4  & \writechord{E}_{5}  \\
        \writechord{D#}_{5} & 3  & & 3  & \writechord{D#}_{5} \\
        \writechord{D}_{5}  & 2  & & 2  & \writechord{D}_{5}  \\
        \writechord{C#}_{5} & 1  & & 1  & \writechord{C#}_{5} \\
        \writechord{C}_{5}  & 0  & & 0  & \writechord{C}_{5}  \\
        \writechord{B}_{4}  & -1 & & -1 & \writechord{B}_{4}  \\
        \writechord{A#}_{4} & -2 & & -2 & \writechord{A#}_{4} \\
        \writechord{A}_{4}  & -3 & & -3 & \writechord{A}_{4}  \\
        \svdots         & \svdots & & \svdots &         \svdots
        \arrow[from=2-2, to=12-4]
        \arrow[from=4-2, to=10-4, color={rgb,255:red,117;green,117;blue,117}, dashed]
        \arrow[from=5-2, to=9-4]
        \arrow[from=7-2, to=7-4, color={rgb,255:red,117;green,117;blue,117}, dashed]
        \arrow[from=9-2, to=5-4]
        \arrow[from=10-2, to=4-4, color={rgb,255:red,117;green,117;blue,117}, dashed]
        \arrow[from=12-2, to=2-4]
      \end{tikzcd}
    };
  \end{tikzpicture}
  \caption{La transformation $A\langle -1,4\rangle :n\mapsto -n + 4$ avec $\alpha =$ \writechord{C}$_5$ et $\beta = 12$}
  \medskip
  \small
  Pour des raisons de lisibilité seules les flèches de la gamme de Do majeur ont été tracées. On notera la passage de l'accord \writechord{Cma} à \writechord{Ami} et de \writechord{Ami} à \writechord{Cma}.
  \label{fig:inversion}
\end{figure}

\begin{figure}
  \includegraphics[width=\columnwidth]{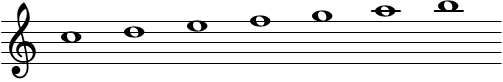}
  \begin{tabularx}{\columnwidth}{ YYYYZ }
    &&$ \mathlarger{\mathlarger{\mathlarger{\mathlarger\Downarrow}}}$ & $A\langle -1,4 \rangle$&
    \end{tabularx}
  \includegraphics[width=\columnwidth]{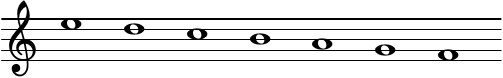}
  \caption{L'image de la gamme de Do majeur par $A\langle -1,4 \rangle$ est un mode de Mi}
  \label{fig:modeE}
\end{figure}

Lorsque $\mu = -1$, les transformations $A \langle -1,\tau\rangle : n\mapsto -n + \tau$ permettent de passer d'une gamme majeure à un mode de Mi et réciproquement (voir Figure \ref{fig:modeE}). La Figure \ref{fig:inversion} illustre la manière dont la triade \writechord{C}$_5$, \writechord{E}$_5$, \writechord{G}$_5$ est transformée par $A\langle -1,4 \rangle$ en la triade \writechord{E}$_5$, \writechord{C}$_5$, \writechord{A}$_4$. De même, l'image de la triade \writechord{A}$_4$, \writechord{C}$_5$, \writechord{E}$_5$  est  \writechord{G}$_5$, \writechord{E}$_5$, \writechord{C}$_5$. Comme les transformations affines préservent les classes de hauteur, on peut affirmer plus généralement que $A \langle -1,4\rangle$ transforme l'accord \writechord{Cma} en \writechord{Ami} et \writechord{Ami} en \writechord{Cma}.

Remarquons de plus qu'en changeant l'ancre,  en choisissant $\alpha =$  \writechord{G}$_4$ par exemple, l'image de l'accord \writechord{Gma} par $A\langle-1,4 \rangle$ sera \writechord{Emi}. La Table \ref{tab:triadesA-14} explicite les images des accords de la gamme de Do majeur par $A\langle -1,4 \rangle$ lorsque $\alpha = $ \writechord{C}$_n$ et des accords de la gamme de Sol majeur lorsque l'on change l'ancre pour $\alpha =$ \writechord{G}$_n$, avec $n$ un entier quelconque.

\begin{table}[htbp]

  \centering 
  \begin{tabular}{cccccccc}
      \writechord{Cma} & $\mapsto$ & \writechord{Ami} & & & \writechord{Gma} & $\mapsto$ & \writechord{Emi}\\
      \writechord{Dmi} & $\mapsto$ & \writechord{Gma} & & & \writechord{Ami} & $\mapsto$ & \writechord{Dma}\\
      \writechord{Emi} & $\mapsto$ & \writechord{Fma} & & & \writechord{Bmi} & $\mapsto$ & \writechord{Cma}\\
      \writechord{Fma} & $\mapsto$ & \writechord{Emi} & & & \writechord{Cma} & $\mapsto$ & \writechord{Bmi}\\
      \writechord{Gma} & $\mapsto$ & \writechord{Dmi} & & & \writechord{Dma} & $\mapsto$ & \writechord{Ami}\\
      \writechord{Ami} & $\mapsto$ & \writechord{Cma} & & & \writechord{Emi} & $\mapsto$ & \writechord{Gma}\\
      \writechord{Bo} & $\mapsto$ & \writechord{Bo} & & & \writechord{F#o} & $\mapsto$ & \writechord{F#o}
  \end{tabular}
  \caption{Images des accords de la gamme de Do majeur par $A\langle -1, 4 \rangle$ pour $\alpha =$ \writechord{C}$_n$ (à gauche) et de la gamme de Sol majeur pour $\alpha =$ \writechord{G}$_n$ (à droite), pour $n\in\mathbb{Z}$}
  \label{tab:triadesA-14}
\end{table}

Ainsi, en se basant sur l'image de l'accord majeur dont la tonique correspond à l'ancre, nous pouvons associer les transformations affines pour lesquelle $\mu = 1$ ou $\mu=-1$ au degré de l'image de cet accord dans la gamme. Par exemple, $A\langle -1,4 \rangle$ peut être associée au degré \writechord{vi} car elle associe le sixième degré mineur à  l'accord majeur dont la tonique est l'ancre (\writechord{Ami} pour \writechord{Cma}, \writechord{Emi} pour \writechord{Gma}, \dots). On obtient ainsi une notation plus intuitive que la notation mathématique, dont la table \ref{tab:degrees} donne un aperçu.

\begin{table}[htbp]
  \centering
  \rowcolors{2}{gray!25}{white}
  \begin{tabular}{ccc}
    \rowcolor{gray!50}
    Degré & Transformation affine\\
    \writechord{I} & $A\langle ~~1, ~~0\rangle$\\
    \writechord{ii} &  $A\langle -1, -3 \rangle$\\
    \writechord{iii} &  $A\langle -1, -1 \rangle$\\
    \writechord{IV} &  $A\langle ~~1,~~ 5 \rangle$\\
    \writechord{V} &  $A\langle ~~ 1, ~~7 \rangle$\\
    \writechord{vi}& $A\langle -1, ~~4\rangle$\\
    \writechord{vii} & $A\langle -1, ~~6 \rangle$\\
  \end{tabular}
  \caption{ Correspondances entre triades d'une gamme majeure et transformations de gamme\label{tab:degrees} }
\end{table}

Remarquons que, de manière générale, les inversions affines se comportent moins bien sur les modes non naturels. Par exemple, $A\langle -1, 4\rangle ($\writechord{G\sharp}$) = $ \writechord{G\sharp} donc l'image de la gamme de La mineur harmonique par $A\langle -1, 4\rangle$ contient les notes \writechord{C}, \writechord{D}, \writechord{E}, \writechord{F}, \writechord{G}, \writechord{G\sharp}, \writechord{B}, ce qui ne correspond pas à un mode standard de la musique tonale. C'est une des faiblesses des transformations affines.

\subsubsection{Transformation vers un mode à transposition limitée}
\begin{figure}[htbp]
  \centering
  \begin{tikzpicture}[baseline= (a).base]

    \node[scale=1] (a) at (0,0){
      \begin{tikzcd}[column sep=0mm, minimum width = 0mm, minimum height=7mm, row sep=0cm]
        \svdots   & \svdots & \hspace{20mm} & \svdots & \svdots \\
        \writechord{G}_{5}  & 7  & & 7  & \writechord{G}_{5}  \\
        \writechord{F#}_{5} & 6  & & 6  & \writechord{F#}_{5} \\
        \writechord{F}_{5}  & 5  & & 5  & \writechord{F}_{5}  \\
        \writechord{E}_{5}  & 4  & & 4  & \writechord{E}_{5}  \\
        \writechord{D#}_{5} & 3  & & 3  & \writechord{D#}_{5} \\
        \writechord{D}_{5}  & 2  & & 2  & \writechord{D}_{5}  \\
        \writechord{C#}_{5} & 1  & & 1  & \writechord{C#}_{5} \\
        \writechord{C}_{5}  & 0  & & 0  & \writechord{C}_{5}  \\
        \writechord{B}_{4}  & -1 & & -1 & \writechord{B}_{4}  \\
        \writechord{A#}_{4} & -2 & & -2 & \writechord{A#}_{4} \\
        \writechord{A}_{4}  & -3 & & -3 & \writechord{A}_{4}  \\
        \svdots         & \svdots & & \svdots &         \svdots
        \arrow[from=5-2, to=1-4, color={rgb,255:red,117;green,117;blue,117}, dotted]
        \arrow[from=6-2, to=3-4]
        \arrow[from=7-2, to=5-4]
       \arrow[from=8-2, to=7-4]
       \arrow[from=9-2, to=9-4]
       \arrow[from=10-2, to=11-4]
       \arrow[from=11-2, to=13-4, color={rgb,255:red,117;green,117;blue,117}, dotted]
    \end{tikzcd}
    };
  \end{tikzpicture}
  \caption{La transformation $A\langle 2,0\rangle :n\mapsto 2n$ permet d'obtenir des gammes apparentées à une gamme par tons.\label{fig:gammepartons}}
\end{figure}
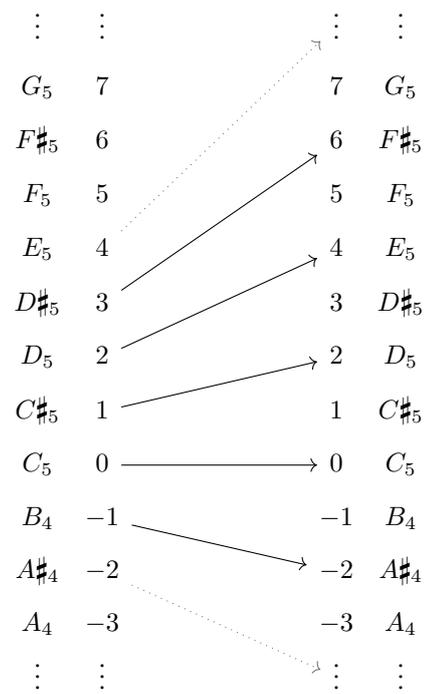

Lorsque $\mu = 2$ ou $\mu = -2$, les transformations affines envoient n'importe quelle gamme vers une gamme apparentée à une gamme par tons (voir Figure \ref{fig:gammepartons}). Les transformations affines peuvent donc permettre de sortir du cadre de la musique tonale.

Contrairement aux inversions et aux transpositions, cette transformation n'est pas bijective : l'image de $A\langle 2,0 \rangle$ con\-tient exactement $6$ classes de hauteurs qui correspondent aux $6$ notes d'une des deux gammes par tons. Il est intéressant de noter que l'image d'une gamme majeure ou mineure naturelle par $A\langle 2,\tau\rangle$ contient les $6$ notes de la gamme par tons (voir Table \ref{tab:minparton}). Ce n'est pas le cas pour la gamme mineure harmonique dont l'image par $A\langle 2,\tau\rangle$ ne contient que $5$ classes de hauteur.

\begin{table}[htbp]
  \centering
  \begin{subtable}[t]{0.24\textwidth}
    \centering 
      \begin{tabular}{ccc}
          \writechord{C} & $\mapsto$ & \writechord{C}\\
          \writechord{D} & $\mapsto$ & \writechord{E}\\
          \writechord{E} & $\mapsto$ & \writechord{G\sharp}\\
          \writechord{F} & $\mapsto$ & \writechord{A\sharp}\\
          \writechord{G} & $\mapsto$ & \writechord{D}\\
          \writechord{A} & $\mapsto$ & \writechord{F\sharp}\\
          \writechord{B} & $\mapsto$ & \writechord{A\sharp}
      \end{tabular}
  \end{subtable}%
  \begin{subtable}[t]{0.24\textwidth}
      \centering 
      \begin{tabular}{ccc}
          \writechord{C} & $\mapsto$ & \writechord{C}\\
          \writechord{D} & $\mapsto$ & \writechord{E}\\
          \writechord{E\flat} & $\mapsto$ & \writechord{F\sharp}\\
          \writechord{F} & $\mapsto$ & \writechord{A\sharp}\\
          \writechord{G} & $\mapsto$ & \writechord{D}\\
          \writechord{A\flat} & $\mapsto$ & \writechord{E}\\
          \writechord{B\flat} & $\mapsto$ & \writechord{G\sharp}
      \end{tabular}
    \end{subtable}
    \caption{Image des gammes de Do majeur (à gauche) et Do mineur naturel (à droite) par $A\langle 2, 0 \rangle$\label{tab:minparton}}
\end{table}

De manière plus générale, lorsque le coefficient modal $\mu$ n'est pas premier avec $\beta$, on obtient un mode à transposition limitée dont les notes sont séparées par un intervalle de $\mu$ demi-tons. Ainsi, pour $\mu = 4$, on obtient un mode composé de $3$ classes d'hauteurs, séparées par des tierces majeures (4 demi-tons).

La Table \ref{tab:classmu} résume les différents types de transformations qu'offrent les transformations affines, ainsi que le nombre de classes de hauteur dans l'image de ces transformations\footnote{Le nombre de classes de hauteur dans l'image de $A\langle \mu, \tau\rangle$ est égal à $\frac{\beta}{\mu\wedge \beta}$ où $\wedge$ dénote le pgcd de deux entiers.}. Les transformations affines bijectives - leur image contient $12$ classes de hauteurs - correspondent  aux automorphismes $F\langle u,j \rangle$ du groupe $T/I$ décrits par \cite{lewin1990klumpenhouwer}.

\begin{table}[htbp]
  \centering
  \rowcolors{2}{gray!25}{white}
  \begin{tabular}{ccc}
    \rowcolor{gray!50}
    $\mu$ & Type de transformation & Classes de hauteur\\
    -1 & Inversions majeur/mineur & 12\\
    0 & Constante & 1\\
    1 & Transpositions & 12 \\
    -2,2 & Gamme par tons & 6 \\
    -3,3 & Tierces mineures &4 \\
    -4,4 & Tierces majeures & 3\\
    -5,5 & $F\langle 7,\tau \rangle$, $F\langle 5,\tau \rangle$& 12 \\
    -6,6 & Tritons & 2\\
    12 & Octaves & 1
  \end{tabular}
  \caption{Classification des transformations affines en fonction de leur coefficient modal $\mu$\label{tab:classmu} }
\end{table}

\subsection{Restriction de l'écart de hauteur}
Jusqu'ici, nous avons présenté les transformations affines en nous concentrant sur leur action sur les classes de hauteur. Dans la pratique, si nous appliquons directement $A\langle -1,4 \rangle$ à la note  \writechord{A}$_4$ qui correspond au La 440Hz et à la note MIDI $69$, on obtient $A\langle -1,4 \rangle(69)  = -65 =$ \writechord{G}$_{-6}$, qui est bien trop grave pour être audible.

Afin de ne pas trop s'éloigner de la tessiture de l'instrument, ou même du spectre auditif, nous allons restreindre l'écart entre la note initiale et son image. Soit $X : \mathbb{Z}\rightarrow \mathbb{Z}$ une transformation de gamme quelconque. Définissons à partir de $X$ une nouvelle transformation $X\langle \delta^-, \delta^+\rangle$ de sorte que $$ - \delta^- \leq X\langle \delta^-, \delta^+\rangle(n) \leq \delta ^+$$ \noindent où $\delta^+$ (resp. $\delta^-$) est l'interval montant (resp. descendant) maximum entre la note initiale et son image.

Soit $\delta = \delta^+ - \delta^-$.  Dans la plupart des cas on souhaite que $\delta = \beta = 12$ , mais il peut être intéressant de choisir par exemple $\delta  = 2\beta$, pour préserver le caractère montant ou descendant d'une ligne mélodique.

Soit $r = |X(n) - n | \mod \beta$ le reste de la division euclidienne de  $|X(n) - n |$ par $\beta$. On pose alors
$$
X\langle \delta^+, \delta^- \rangle : n \mapsto \begin{cases}
  n + r & \text{si $r \leq \delta^+$}\\
  n + r - \delta & \text{sinon}
\end{cases}
$$

Nous pouvons maintenant appliquer cette restriction de l'intervalle de hauteur à nos fonctions affines. On obtient alors un sous-ensemble de transformations de gamme de la forme $A\langle \mu, \tau, \delta^-, \delta^+\rangle$. Ce sont exactement ces transformations, combinées avec les tempéraments $X\langle \alpha,12\rangle$, qui sont implémentées dans LiveScaler.

\section{Implémentation des transformations affines : LiveScaler}

Nous allons à présent nous intéresser à l'implémentation des transformations affines que nous venons de présenter afin de pouvoir les appliquer en live à l'ensemble des flux MIDI qui composent le morceau joué.

\subsection{Architecture de LiveScaler}

\begin{figure}[htbp]
  \centering
	\includegraphics[width=\columnwidth]{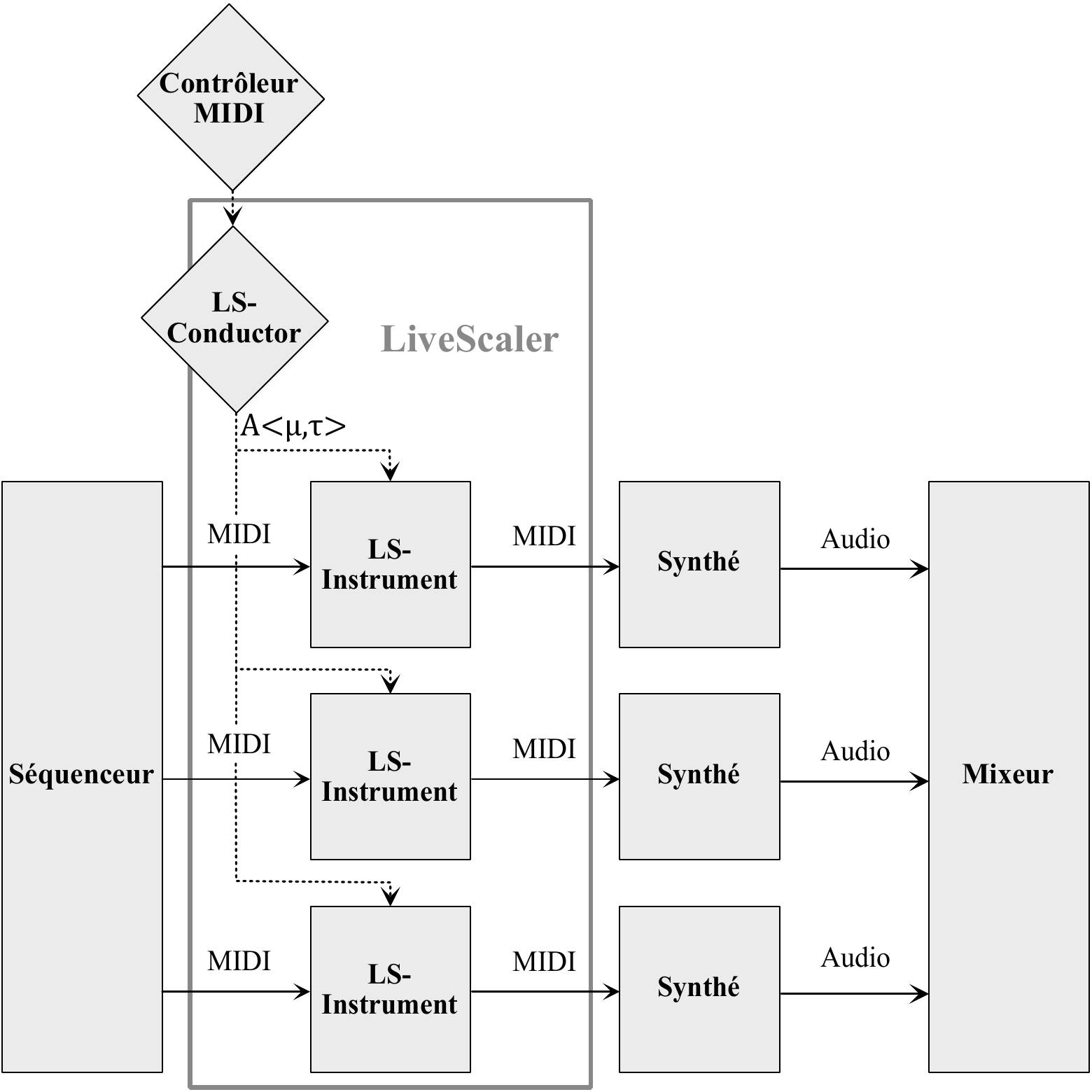}
  \caption{Architecture de Live Scaler\label{fig:archi}}
\end{figure}

LiveScaler fonctionne à la manière d'un orchestre dont le  DJ serait le chef. Chaque piste MIDI contenant un instrument virtuel (synthétiseur, sampleur, etc.) est un instrumentiste de l'orchestre. On souhaite que sur un geste du DJ, chaque instrument virtuel interprète différemment sa partition, c'est à dire le flux MIDI qu'il reçoît. Dans le cadre de LiveScaler, le DJ envoie les paramètres d'une transformation affine à tous les instruments simultanément et ceux-ci doivent appliquer cette transformation dès qu'ils la reçoivent.

L'implémentation de LiveScaler est donc séparée en deux outils interdépendants :
\begin{enumerate}
  \item une interface (appelée \emph{Conductor}, en référence à l'analogie avec l'orchestre) qui récupère les entrées de l'utilisateur (ici le DJ) et les convertit en paramètres d'une transformation affine puis envoie ces paramètres à toutes les instances de \emph{Instrument}.
  \item un plug-in MIDI appelé \emph{Instrument} qui transforme le flux MIDI entrant en appliquant à toutes les notes la transformation affine dont les paramètres ont été reçus de \emph{Conductor}.
\end{enumerate}
La Figure \ref{fig:archi} illustre l'architecture globale de LiveScaler.

On distingue également les paramètres \emph{locaux}, qui sont propres à chaque instrument, et les paramètres \emph{globaux}, qui sont reçus du chef d'orchestre et donc communs à tous les instruments. Dans le cadre des transformations affines, les paramètres $\mu$, $\tau$, $\alpha$  et $\beta$ sont globaux, il correspondent dans une certaine mesure à l'harmonie actuelle du morceau. Quant à $\delta^-$ et $\delta^+$, ils sont locaux et peuvent être adaptés à la tessiture de l'instrument.
\begin{figure*}[h]
  \centering
  \begin{subfigure}{0.83\textwidth}
    \includegraphics{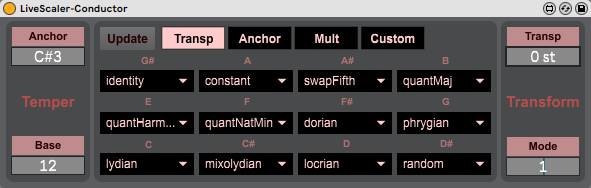}
  \end{subfigure}
  \begin{subfigure}{0.15\textwidth}
    \includegraphics{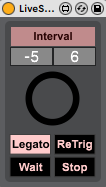}
  \end{subfigure}

  \caption{Interface graphique de LiveScaler (\emph{Conductor} à gauche  et  \emph{Instrument} à droite) }
  \label{fig:LiveScalerUI}
\end{figure*}

\subsection{Quand appliquer les transformations ? }
Lorsque \emph{Instrument} reçoit la commande d'appliquer une nouvelle transformation de gamme, celle-ci est sensée prendre effet immédiatement et être appliquée à toutes les notes reçues jusqu'au prochain changement de gamme. Lorsque l'instrument n'est pas en train de jouer, cela ne pose aucune difficulté : il appliquera la nouvelle transformation au prochain message MIDI qu'il recevra. Il se peut en revanche que l'instrument soit déjà en train de jouer une note. LiveScaler propose $4$ manières de réagir dans une telle situation :

\begin{enumerate}
  \item \emph{Stop} : toutes les notes en train d'être jouées sont instantanément arrêtées en envoyant un message Note-off pour chaque note en cours. L'instrument reprendra son jeu, en appliquant la nouvelle transformation, au prochain message MIDI qu'il recevra. Cette option est particulièrement adaptée aux instruments dont la durée des notes est courte.
  \item \emph{Legato} : chaque note en cours est stoppée et instantanément remplacée par son image par la nouvelle transformation. Si l'instrument virtuel est paramétré sur \emph{Legato}, alors les changements de gamme déclencheront des legatos.
  \item \emph{ReTrigger} : agit sur le même principe que \emph{Legato} à la différence  qu'un court délai est introduit entre la fin de la note en cours et la note transformée, forçant une nouvelle attaque, même si l'instrument virtuel est en mode legato.
  \item \emph{Wait} : les notes en cours continuent d'être jouées telles quelles. Si elles ne se sont pas arrêtées avant, elles seront stoppée lorsque la prochaine note sera jouée, à partir de laquelle la nouvelle transformation prendra effet.
\end{enumerate}

Le choix entre ces quatre modes se fait de manière locale, deux instances de \emph{Instrument} pourront donc réagir différemment.

\subsection{Implémentation avec Max for Live}

L'objectif principal de l'implémentation proposée était de pouvoir expérimenter le plus rapidement possible sur les transformations en tant que musicienne. Max for Live est une intégration du  langage de progammation graphique Max MSP à Ableton Live. On peut aisément communiquer avec les différences instances du logiciel, ce qui permet dans LiveScaler à \emph{Conductor} de contrôler les \emph{Instruments} avec une faible latence\footnote{En moyenne, LiveScaler introduit une latence inférieure à $1$ ms.}. La station audionumérique Ableton Live étant particulièrement populaire pour composer et produire de la musique dans le milieu de l'EDM, c'est donc naturellement que nous avons choisi Max for Live pour une première implémentation.

La Figure \ref{fig:LiveScalerUI} montre l'interface graphique de LiveScaler. L'implémentation actuelle permet d'appliquer les transformations affines et les transformations périodiques sur un intervalle. Pour ces dernières, il suffit de renseigner manuellement l'image de chaque note de l'intervalle considéré dans un fichier pour y avoir ensuite accès via LiveScaler.

\section{Escape : une performance avec LiveScaler }
Afin de tester la pertinence des transformations proposées, j'ai utilisé LiveScaler pour créer \emph{Escape}, une performance live de trance psychédélique. J'ai notamment pu proposer cette performance en publique pour la soutenance de mon mémoire de master en Septembre 2019. Dans cette section, j'expliquerai la manière dont j'ai procédé pour cette mise en pratique et je donnerai mes impressions subjectives en tant que musicienne et utilisatrice de LiveScaler.

\subsection{Contrôle live dans Escape}
\begin{figure}[htbp]
  \centering
  \includegraphics[width=\columnwidth]{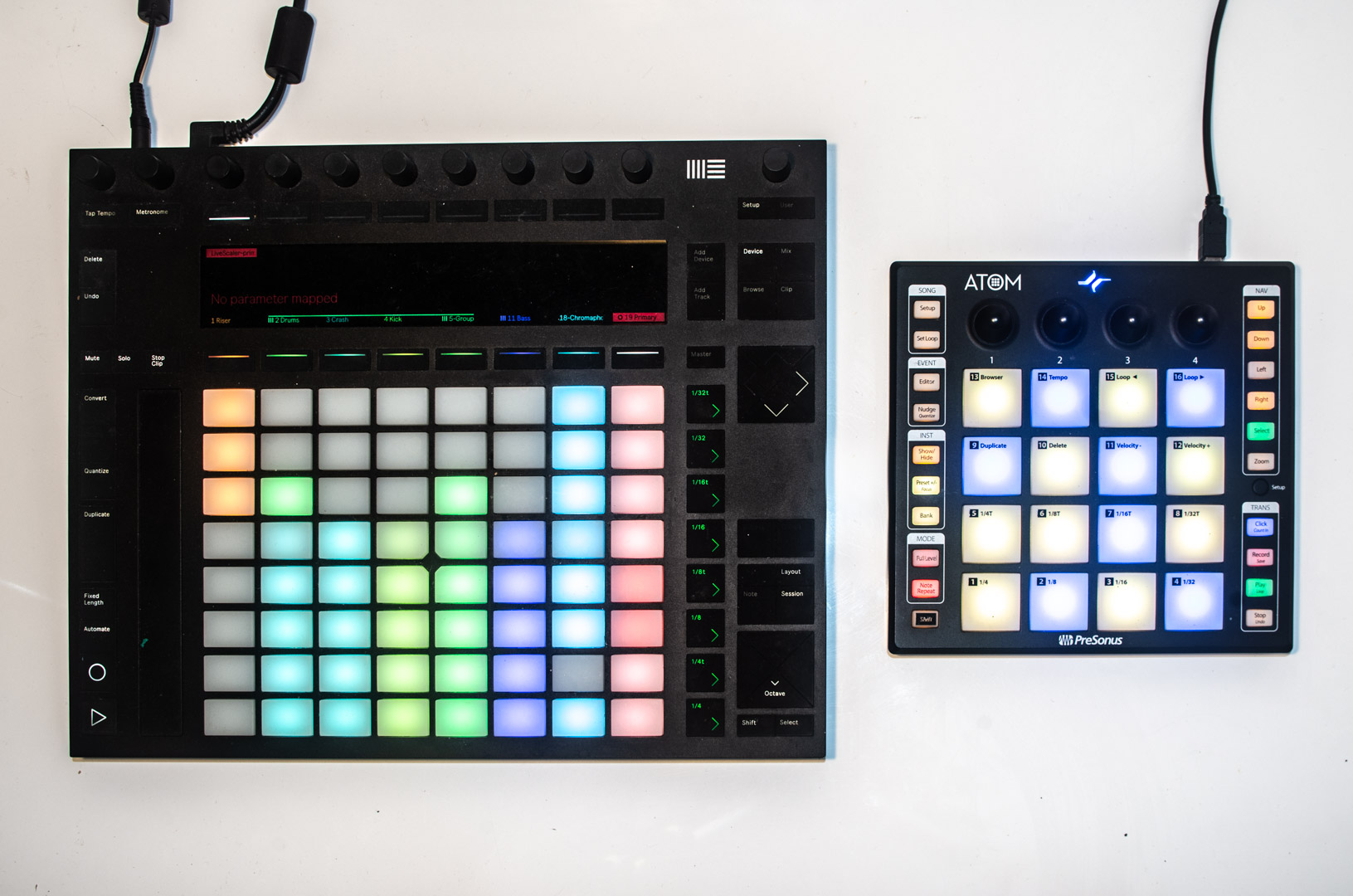}
  \caption{Contrôleurs MIDI utilisés pour la performance live : à gauche le Push 2 par Ableton (contrôle de la structure du morceau) et à droite ATOM par Presonus (contrôle de l'harmonie du morceau).\label{fig:controleurs}}
\end{figure}
Pour \emph{Escape}, j'utilise deux contrôleurs MIDI distincts (voir Figure \ref{fig:controleurs}) :
\begin{itemize}
  \item le Push 2 par Ableton, qui est conçu spécialement pour contrôler Ableton Live. Il me permet de recréer en live la structure d'Escape en déclenchant ses différentes parties.
  \item l'ATOM de Presonus, composé d'une grille de $4\times 4$ touches qui me permettent de déclencher les transformations de LiveScaler.
\end{itemize}

\begin{figure}[htbp]
  \centering
  \includegraphics[width=\columnwidth]{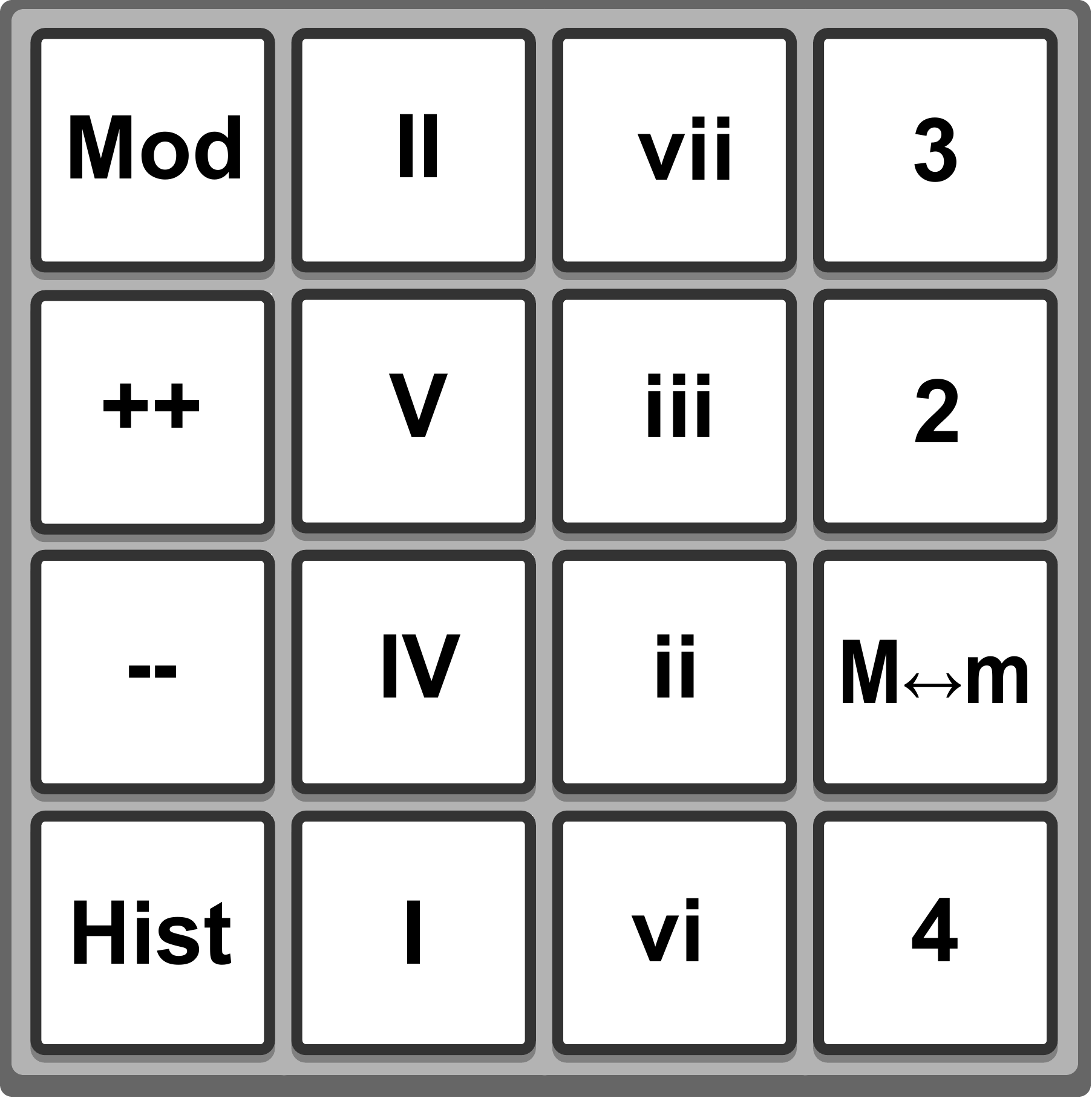}
  \caption{Mapping de LiveScaler sur un contrôleur MIDI à $4\times 4$ touches\label{fig:mapping-ATOM}}
\end{figure}

La figure \ref{fig:mapping-ATOM} illustre la manière (le plus souvent appelée \emph{mapping}) dont les touches du contrôleur sont associées aux transformations de LiveScaler.

Voici le détail de l'action des différentes touches :
\begin{itemize}
  \item \LSI, \LSvi, \LSIV, \LSII, \LSV, \LSiii, \LSII, \LSvii :  les deux colonnes centrales déclenchent instantanément les transformations décrites précédemment. Elles sont organisées par relatives mineures/majeures.
  \item  \texttt{Hist} : LiveScaler garde en mémoire un court historique des transformation précédemment appliquées. Appuyer sur \texttt{Hist} permet de déclencher une des transformations de cet historique. Des combinaisons de la touche \texttt{Hist} et des touches \texttt{Hist}, \LSMm, \LStwo, \LSthree, et \LSfour permettent de naviguer dans cet historique \footnote{Pour plus de détail, se référer au manuel de LiveScaler}.
  \item \LSpp, \LSmm : applique  $\tau = \tau + 1$ (resp. $\tau = \tau - 1$). En pratique, combiner la touche $++$ (resp. $--$) avec une des transformations des colonnes centrales, on transpose cette transformation d'un demi-ton vers le haut (resp. vers le bas).
  \item  \LSMm : applique $\tau = \tau + 5\mu $ et $\mu = -\mu$.  En pratique, combiner \LSMm avec une des transformations centrales permet de passer d'une transposition à une inversion et réciproquement : combiner \LSMm avec \LSI $~$ (resp. \LSii, \LSiii, \LSIV, \LSV, \LSvi, \LSvii) donnera la transformation \LSi $~$ (resp.  \LSII, \LSIII, \LSiv, \LSv, \LSVI, \LSVII) et réciproquement.
  \item \LStwo, \LSthree, \LSfour : applique $\mu = 2\mu$ (resp. $\mu = 3\mu$, $\mu = 4\mu$). On obtient ainsi les modes à transposition limitée décrits précédemment.
  \item \LSMod : en combinant \LSMod avec une des transformations, on indique à LiveScaler qu'on souhaite moduler l'harmonie de notre morceau vers cette nouvelle gamme.
\end{itemize}

Ainsi, on pourrait imaginer harmoniser en live une instrumentation jouant de manière répétée sur l'accord \writechord{C}. Par exemple, si on veut reproduire la suite d'accord de la chanson \emph{Summer Nights} de la comédie musicale \emph{Grease} \footnote{Si, comme pour moi, cette chanson à tendance à rester dans votre tête, je suis (presque) désolée} en commençant par indiquer à LiveScaler qu'on est dans une tonalité de Ré majeur (\LSMod + \LSII). Puis, une fois l'instrumentation lancée, on appuiera successivement tous les deux temps sur  \LSI - \LSIV - \LSV - \LSIV.

Puis, arrivés au moment tant attendu de la modulation d'un demi-ton vers le haut, on indique à LiveScaler

\begin{center}
  \LSI $~$-  \LSIV $~$-  \LSV $~$-  (\LSpp + \LSMm +  \LSvi) $~$-  (\LSMod + \LSpp + \LSI)
\end{center}

\noindent pour repartir joyeusement sur \LSI - \LSIV - \LSV - \LSIV, mais cette fois dans une tonalité de Mi bémol majeur. On obtiendrait ainsi la progression harmonique suivante :

\begin{center}
  \dots \hspace{4pt}- \writechord{D} - \writechord{G} - \writechord{A} - \writechord{G} - \writechord{D} - \writechord{G} - \writechord{A} - \writechord{Bb} - \writechord{Eb} - \writechord{Ab}  - \writechord{Bb} - \writechord{Ab} - \dots
\end{center}

\subsection{Processus de composition}

J'ai composé \emph{Escape} dans le but de le jouer avec LiveScaler. C'est un morceau de trance psychédélique (psytrance)\footnote{La trance psychédélique est souvent  appelée \emph{psytrance}, le lectorat curieux pourra écouter par exemple l'album \emph{The Gathering} (1999) du duo israëlien \emph{Infected Mushroom}.} composé  :
\begin{itemize}
  \item d'une mélodie minimaliste durant 2 mesures jouée par un synthétiseur  dont le son se rapproche d'un métallophone éthéré
  \item d'une \emph{rolling bass} classique devenue une des signatures de la \emph{psytrance}, et de plusieurs autres basses sur une unique note pédale produisant une ligne de basse riche dans sa texture et son timbre
  \item d'une rythmique séquencée à l'avance indépendante de LiveScaler (pistes audio)
  \item de quelques effets sonores (\emph{risers}, \emph{downshifters} \dots) eux aussi typiques de la \emph{psytrance}, que je déclenche ponctuellement à l'aide du Push.
\end{itemize}

L'objectif était d'une part de partir d'un morceau simple, basé sur un unique accord (ici \writechord{Am}) et d'enrichir son harmonie et sa ligne mélodique avec LiveScaler; d'autre part de proposer un morceau typique d'un genre de musique électronique populaire très codifié (ici la \emph{psytrance}). Le choix du genre n'est pas anodin :  je souhaitais utiliser un outil de contrôle live expérimental pour un morceau de musique qui, lui, n'a rien d'expérimental. Pour moi, il s'agit plus avec LiveScaler d'explorer de nouvelles modalités live que de nouveaux horizons musicaux.

Pour autant, LiveScaler peut tout à fait intervenir dans le processus de composition. On peut par exemple l'utiliser pour trouver des variations sur une mélodie ou une arpège en expérimentant avec les différents changements de gamme, puis consolider le MIDI une fois qu'on a trouvé une idée satisfaisante. On obtient alors un processus créatif incrémental et prône à la sérendipité partant d'une mélodie ou d'une progression d'accords simple qu'on enrichit ensuite avec LiveScaler.

\section{Travaux connexes}

Les transformations affines sont directement inspirées de la théorie transformationnelle  (pour une introduction généraliste du point de vue mathématique, voir \cite{andreatta2008calcul}, et du point de vue musicologique, lire \cite{andreatta2014introduction}). La théorie transformationnelle prend ses racines dans la \emph{Set-Theory}, qui se concentre sur la notion de classe de hauteur, c'est à dire un ensemble de notes identiques à l'octave près \cite{forte1973structure}. Elle propose une approche plus algébrique que la \emph{Set-Theory}, en se concentrant, entre autre sur la notion de transformation entre ensembles de classes de hauteurs \cite{lewin1987generalized}.

Les transformations affines sont particulièrement proche des automorphismes du groupe T/I proposés par David Lewin \cite{lewin1990klumpenhouwer}. L'unique différence avec ceux-ci est que les transformations affines ne sont pas nécessairement bijectives et autorisent donc un coefficient modal qui ne soit pas nécessairement premier avec $12$. Les transformations affines sont donc  une vision plus appliquée (l'aspect algébrique est passé sous silence) des automorphismes proposés par Lewin et Klumpenhouwer, tout en proposant quelques transformations supplémentaires sortant du système tonal.

Plusieurs outils s'appuient plus ou moins explicitement sur les représentations de la théorie transformationnelle, en particulier OpenMusic, qui propose une aide à la composition directement inspirée de celle-ci \cite{andreatta2003implementing}, \cite{andreatta2003formalisation}. Depuis une dizaine d'années, OpenMusic essaie de concilier l'approche hors du temps (approche guidée par les demandes) de l'aide à la composition avec l'approche temps-réel propre à la performance \cite{bresson2014reactive}, \cite{bresson2017next}. Plus récemment, Bach propose lui aussi cette approche hybride mais en partant de Max MSP, un language de programmation fondamentalement temps-réel et guidé par les données \cite{agostini2021programming}.

Les outils évoqués ci-dessus offrent une grande flexibilité pour appliquer des transformations potentiellement bien plus sophistiquées que les transformations affines, et sont \emph{a priori} tous capables de le faire en live. Pour autant, cela demanderait un grand travail préparatoire de programmation et de \emph{mapping} avant d'arriver à un résultat fluide. C'est exactement ce travail qui est fait par LiveScaler, mais sur un nombre restreint de transformations, ici jugées pertinentes. LiveScaler sacrifie donc la flexibilité dans le choix des transformations au profit d'une utilisation immédiate, et sans connaissances de progammation requises, pour faire de la musique live.

\paragraph*{}

Une approche intermédiaire offrant plus de flexibilité, mais moins d'immédiateté est celle du \emph{live coding}. Pendant une performance de \emph{live coding}, le musicien utilise un langage de programmation dédié pour coder en live un morceau de musique \cite{blackwell2022live}. En particulier le langage Tidal \cite{mclean2010tidal} permet de manipuler et transformer des motifs en live. Il reprend notamment les transformations de Laurie Spiegel \cite{spiegel1981manipulations} et plus particulièrement les transpositions et inversions (qui correspondent aux coefficients modaux $1$ et $-1$ dans le paradigme des transformations affines proposé ici).

Si une des revendications initialement associées à la pratique du \emph{live coding} était de s'affranchir des contraintes et rigidités des stations audionumériques telle qu'Ableton Live \cite{collins2003live}, l'ajout de langages de scripting  ainsi que la posibilité de contrôler les stations numériques \footnote{On peut par exemple piloter  Ableton Live et FL Studio avec Python, Logic et Bitwig avec JavaScript.} avec des langages de programmation graphique haut niveau \footnote{On trouve dans les stations audionumériques commerciales de plus en plus de langages de "\emph{patching}" permettant de contrôler le logiciel ou de créer des plugins audio de manière modulaire : voir par exemple Max for Live pour Ableton Live, FL FlowStone pour FL Studio, The Grid pour Bitwig.} semble avoir développé leur usage dans les pratiques de musique algorithmique live \cite{collins2014algorave}. Ces pratiques, dans lesquelles s'incrit le présent article, permettent de combiner d'une part la flexibilité et la liberté qu'offrent un langage de programmation et d'autre part l'accès aux outils de production commerciaux utilisés par l'industrie de la musique. C'est particulièrement important dans le cadre de l'EDM, qui utilise intensivement ces outils de production sophistiqués \cite{fraser2012spaces}.

\paragraph*{}
L'idée d'utiliser un morceau composé au préalable comme matière première à laquelle on applique des transformations est particulièrement développée, et théorisée par Louis Bigo et Darrell Conklin \cite{bigo2016viewpoint}. En analysant au préalable l'harmonie d'un morceau, ils proposent ainsi de la transformer ensuite afin d'obtenir une variation du morceau. LiveScaler se démarque de cette approche par deux principaux aspects : LiveScaler ne nécessite aucune analyse préalable et se concentre sur le live. Une fois de plus la contrepartie est une moins grande flexibilité dans le choix des transformations.

Le logiciel EmoteControl \cite{micallef2021emotecontrol} est sans doute celui qui se rapproche le plus de LiveScaler. Il permet un macrocontrôle en live de paramètres tels que le tempo, l'articulation, la hauteur de note, etc. En particulier, il propose une inversion du mode (qui correspond au coefficient modal $-1$).  LiveScaler offre une bien plus grande variété de transformations de gammes et de flexibilité sur la performance live. Une collaboration serait ici particulièrement intéressante : le contrôle de l'articulation ou du timbre seraient particulièrement pertinents à ajouter à LiveScaler.

\section{Conclusion}
Nous avons présenté LiveScaler, qui propose de nouvelles modalités pour la musique électronique live, ansi qu'un jeu de transformations MIDI : les transformations affines.  Live\-Scaler permet d'appliquer ces transformations affines en live. En particulier, cet outil peut être utilisé dans le contexte de l'EDM, proposant ainsi une nouvelle alternative ou un complément pour la performance live en musique électronique.

Plusieurs améliorations techniques pourraient être proposées pour améliorer LiveScaler. En particulier, il serait pertinent de le rendre compatible avec n'importe quel DAW, et pas seulement Ableton Live. Une solution sera de  développer un plugin VST ou LV2 pour LiveScaler. De plus, le protocole MIDI étant contraignant, rendre LiveScaler compatible avec un protocole plus flexible tel que OSC \cite{wright2005open}, MPE (Midi Polyphonic Expression), Midi 2.0 ou encore  MP \cite{goudard2017mapping}, qui serait particulièrement adapté à cette application.

Bien que LiveScaler propose déjà la possibilité d'ajouter manuellement des transformations de gamme périodiques sur un intervalle, ce mécanisme est laborieux et mérite d'être amélioré. Utiliser des transformations sur un espace diatonique serait également intéressant et soulèverait la difficulté de connaître la tonalité dans laquelle on se trouve. Enfin, nous aimerions pouvoir agir sur d'autres paramètres que la hauteur des notes, en particulier le rythme, le timbre, ou même contrôler simultanément des transformations musicales et vidéos. C'est sur ces axes que seront concentrées nos recherches futures, tout en restant sur le même paradigme de performance live que celui proposé ici.

\subsection{Remerciements}
Je tenais à remercier ici David Janin et Martin Laliberté, mes encadrants de thèse, pour leur relecture et leurs nombreux conseils. Merci également à Chloé Lavrat, pour avoir pris le temps de me lire et pour nos nombreuses discussions sur le sujet, toujours très inspirantes.

\newpage


\end{document}